\documentclass[aps,prb,reprint]{revtex4-1}
\usepackage{blindtext}
\usepackage{epsfig}
\usepackage{color}
\usepackage{amsmath}
\usepackage{amsfonts}
\usepackage{amssymb}
\usepackage{graphicx}%
\usepackage{url}
\usepackage[breaklinks]{hyperref}
\usepackage[english]{babel}
\usepackage{esvect}
\usepackage[normalem]{ulem}
\usepackage[inline]{enumitem}

\hypersetup{colorlinks,citecolor=blue,filecolor=blue,linkcolor=blue,urlcolor=blue}

\begin{document}
\title{Real-Time Ramsey Interferometry in Fractional Quantum Hall States}

\author{Tal Goren and Karyn Le Hur}
\affiliation{Centre de Physique Th\'{e}orique, \'{E}cole Polytechnique, CNRS, Universit\'{e} Paris-Saclay,  91128 Palaiseau, France}

\begin{abstract}
We study the time-dependent dynamical phase of fractional quasiparticles in quantum Hall states using Ramsey interferometry.
A Ramsey two-pulse voltage modulation generates an interference term in the current noise of a quantum point contact oscillating with the dynamical phase. We show that the interference pattern probes the Green's function of the fractional quasi-particles in the time domain and reveals the fractional charge. We address both the case of a point-like tunnel junction and the case of momentum-resolved tunneling which also provides information on the speed of those quasiparticles propagating along the edge.  For the $5/2$ fractional quantum Hall case, the Ramsey signal can differentiate between
the Pfaffian and anti-Pfaffian states. 
\end{abstract}

\date{\today}
\maketitle


One of the main attributes of the fractional quantum Hall effect is the occurrence of excitations with fractional charge and statistics. The edge states of such a quantum Hall system are described by a chiral Luttinger liquid effective theory which illustrates the emergence of fractional quasi-particles due to electron-electron interactions in quantum Hall states.  \cite{stone1994,wen2004book}


Ramsey interferometry has been recently suggested to probe the dynamics of mesoscopic electron systems \cite{Tal2016} and strongly correlated systems such as  one-dimensional quasicondensates \cite{Kitagawa2010}, spin ensembles \cite{Knap2013} and Rydberg atoms \cite{Ruseckas2017}. Here, we study Ramsey interferometry in the Abelian and 5/2 non-Abelian fractional quantum Hall states at the edges, through the quantum Hall bar geometry.

\begin{figure}
\begin{center}
\includegraphics[width=0.45\textwidth]{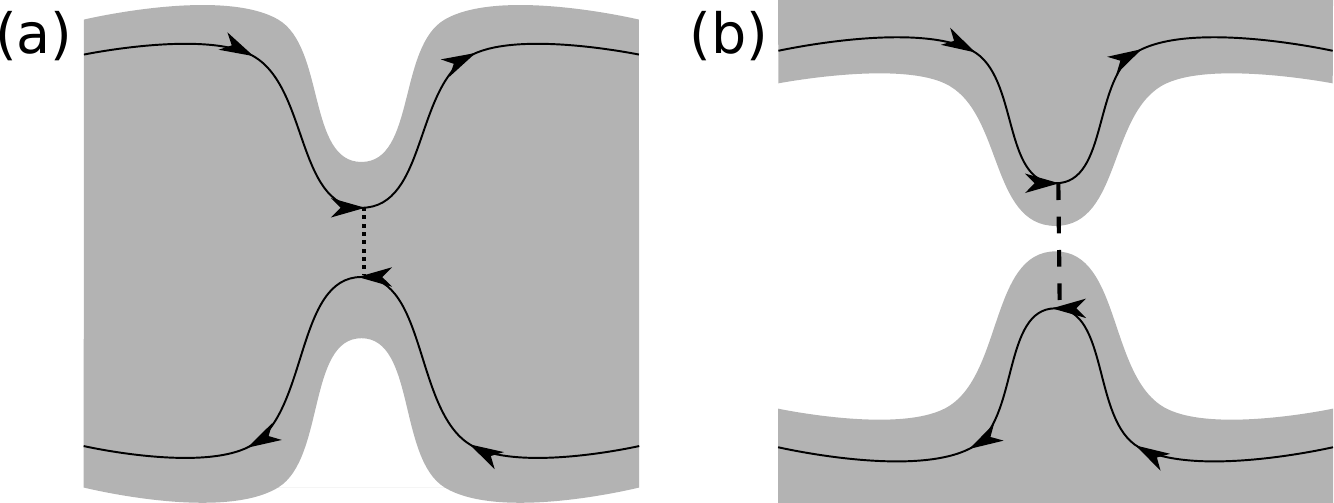}
\end{center}
\vskip -0.5cm
\protect\caption[QPC Geometries]{  \it Illustration of the two geometries of a QPC. The quantum Hall fluid is marked in grey. The black lines indicate the chiral edge states.  a) Tunneling between the two edges of a Hall bar: the transport is driven by Laughlin quasiparticles. The tunneling particles are the fractional quasi-particles. b) Tunneling between two separated Hall droplets. The tunneling particles here are electrons.
\label{fig:QPC}
}
\end{figure}

A small constriction in the Hall bar induces a coupling between the counter-propagating edge states (see Fig. \ref{fig:QPC}). Such a quantum point contact (QPC) leads to tunneling of particles from one edge to the other. Applying a Ramsey voltage modulation composed of two short pulses involves tunneling of particles during the voltage pulses. Ramsey interference is the result of the interference between a particle tunneling during the first pulse and a particle tunneling during the second pulse and it oscillates with the time delay between the pulses which we denote by $t_0$. Therefore the Ramsey protocol allows for a time resolved measurement of the dynamical phase and Green's function of the tunneling particles by scanning the time delay $t_0$ between the voltage pulses.

We consider Laughlin states  with filling factor $\nu=\frac{1}{2n+1}$ where $n$ is an integer \cite{Laughlin} although our method can be easily adapted to other quantum Hall states. The fractional quasi-particle  with charge $q=e\nu$ creation operator  is expressed by the bosonic chiral fields $\phi_{R/L}\left(x\right)$ \cite{wen2004book,FisherGlazman1997,giamarchi2004}
\begin{equation}
\psi_{f,R/L}^{\dagger}\left(x\right)\sim e^{i\phi_{R/L}\left(x\right)}
\end{equation}
where $R/L$ denote the right/left moving particles on different edges.
An electron is composed of $\nu^{-1}$ fractional quasi-particles, thus its creation operator is given by
\begin{equation}
\psi_{e,R/L}^{\dagger}\left(x\right)\sim e^{i\frac{\phi_{R,L}\left(x\right)}{\nu}}.
\end{equation}
Here, $L$ and $R$ refer to left/right moving electrons on different edges. The commutation relations
\begin{equation}\begin{split}
\left[\phi_{L}\left(x_{1}\right),\phi_{R}\left(x_{2}\right)\right]&=0\\
\left[\phi_{R/L}\left(x_{1}\right),\phi_{R/L}\left(x_{2}\right)\right]&=\pm i\pi \nu\text{sign}\left(x_{1}-x_{2}\right)
\end{split}\end{equation}
ensure the correct anti-commutation relation for the electrons and lead to a fractional statistics for the quasi-particles
\begin{equation}
\psi_{f,R/L}\left(x\right)\psi_{f,R/L}\left(x'\right)=e^{i\pi\nu}\psi_{f,R/L}\left(x'\right)\psi_{f,R/L}\left(x\right).
\end{equation}

A quantum point contact can be designed in two geometries represented in Fig. \ref{fig:QPC}. A constriction in the Hall bar leads to tunneling of fractional quasi-particles through the bulk of the Hall bar (see Fig. \ref{fig:QPC}a). Alternatively, points on two different quantum Hall droplets can be pinched  close to each other (see Fig. \ref{fig:QPC}b). In such a case only electrons or holes can tunnel. 
Below, we study the Ramsey protocol for these two scenarios.

%




The tunneling of particles from one edge to another via a QPC located at $x=0$ is described by the Hamiltonian:\cite{wen2004book,FisherGlazman1997,giamarchi2004}
\begin{equation}
H_{QPC}=De^{i\varphi\left(t\right)}\psi_{L}^{\dagger}\left(x=0\right)\psi_{R}\left(x=0\right)+h.c.
\end{equation}
where the dynamical phase $\varphi(t)$  is defined by
\begin{equation}
\varphi\left(t\right)\equiv q\int^{t}dt'V\left(t'\right),
\end{equation}
where $V(t)$ is the voltage drop across the QPC and $q$ is the charge of the tunneling particle.  The dynamical phase factor encodes the charge of the tunneling particle ($\psi=\psi_f$ or $\psi=\psi_e$).
Hereafter, we fix the Planck constant to $\hbar=h/(2\pi)=1$. 


Quantum interferences between tunneling events at different times can be revealed by the current noise.
 Hence we will derive the effect of a time-dependent voltage modulation on the current noise. The current (operator) between the two edges flowing through the QPC takes the form
\begin{equation}
I_{QPC}=iqDe^{i\varphi\left(t\right)}\psi_{L}^{\dagger}\left(x=0\right)\psi_{R}\left(x=0\right)+h.c.
\end{equation}
Since we consider time-dependent voltage modulations we apply the Keldysh formalism which is suitable to describe non-equilibrium systems \cite{kamenev2011}. To lowest order in perturbation theory, the current-current correlation function symmetrized on the 
two branches of the Keldysh contour is
\begin{equation}
M_{2}(t_1,t_2)=\frac{1}{2^{2}}\sum_{\eta_{1}\eta_{2}=\pm}\left\langle I_{QPC}\left(t_{1}^{\eta_{1}}\right)I_{QPC}\left(t_{2}^{\eta_{2}}\right)\right\rangle,
\end{equation}
where $\eta_{1/2}$ are Keldysh indices and the expectation value $\langle\rangle$ is taken with respect to a system without a QPC.
The current involves an operator at $x=0$ only, thus to shorten the notations hereafter we suppress the position of the operators.
We find the current-current correlation function to second order in $D$
\begin{multline} \label{eq:M_2}
M_{2}\left(t_{1},t_{2}\right)=\frac{1}{4}q^{2}D^{2}\sum_{\eta_{1}\eta_{2}=\pm} \\
\bigg[\langle\psi_L^\dagger(t_1^{\eta_1})\psi_L(t_2^{\eta_2}) \rangle \langle\psi_R(t_1^{\eta_1})\psi_R^\dagger(t_2^{\eta_2}) \rangle e^{i\varphi(t_{1})}e^{-i\varphi(t_{2})} \\
+ \langle\psi_L(t_1^{\eta_1})\psi_L^\dagger(t_2^{\eta_2}) \rangle \langle\psi_R^\dagger(t_1^{\eta_1})\psi_R(t_2^{\eta_2}) \rangle e^{-i\varphi(t_{1})}e^{i\varphi(t_{2})} \bigg].
\end{multline}

The current noise is defined as the Fourier transform of the current-current correlation function
\begin{equation} \label{eq:Noise_Def}
S(\omega)=\int_{-\infty}^{\infty}dt_1dt_2M_2(t_1,t_2)e^{i\omega(t_1-t_2)}.
\end{equation}
For a $dc$ voltage Eq. \eqref{eq:Noise_Def} results in the well known Schottky formula \cite{Kane1994,Fendley1995,Chamon1995}. Now, we shall also include the effect of time-dependent voltage modulations. 
Within the stationary phase approximation the main contributions to the integral in Eq. \eqref{eq:Noise_Def} come from times where the argument of the imaginary exponent is stationary. The stationary conditions for the first term in Eq. \eqref{eq:M_2} are
\begin{equation}
\dot{\varphi}\left(t_{1}\right)+\omega=0\quad-\dot{\varphi}\left(t_{2}\right)-\omega=0
\end{equation}
whereas for the second term in Eq. \eqref{eq:M_2} they are
\begin{equation}
-\dot{\varphi}\left(t_{1}\right)+\omega=0 \quad\dot{\varphi}\left(t_{2}\right)-\omega=0.
\end{equation}

For simplicity we consider a positive voltage modulation $\varphi(t)\geq 0$. Hence for positive frequencies, the main contributions to the noise are solely from the second term of Eq. \eqref{eq:M_2} at times $t_1,t_2$ satisfying 
$\dot{\varphi}(t_1)=\dot{\varphi}(t_2)$. The double-integral can be approximated by its integrand at equal voltage times
\begin{equation}\label{eq:V_cond}
V(t_1)=V(t_2)=\omega,
\end{equation}
resulting in
\begin{multline}
S(\omega>0)\sim\frac{1}{4}q^{2}D^{2} \\
\times\sum_{\{t_1,t_2\vert V(t_1)=V(t_2)\}}e^{-i\varphi(t_{1})+i\omega t_1}e^{i\varphi(t_{2})-i\omega t_2} \\
\times\sum_{\eta_{1}\eta_{2}=\pm}G_L^{\eta_1\eta_2}(t_1-t_2) G_R^{\eta_1\eta_2}(t_1-t_2),
\end{multline}
where we have defined the Green's functions of the tunneling particles
\begin{equation}\begin{split}
G_L^{\eta_1\eta_2}(t_1-t_2)=\langle\psi_L(t_1^{\eta_1})\psi_L^\dagger(t_2^{\eta_2}) \rangle \\
G_R^{\eta_1\eta_2}(t_1-t_2)=\langle\psi_R^\dagger(t_1^{\eta_1})\psi_R(t_2^{\eta_2}) \rangle.
\end{split}\end{equation}

Next, we consider the Ramsey voltage modulation composed of two short pulses separated by the delay time $t_0$ on top of a constant voltage
\begin{equation}
V(t)=V_0(t)+V_0(t-t_0)+V_{dc}.
\end{equation}
For pulse durations much smaller than the delay time $t_0$ we can approximate the pulse by a Dirac delta function $qV_0(t)\approx\varphi_0\delta(t)$.
Therefore, there  are four possibilities to satisfy the stationary condition \eqref{eq:V_cond} for the Ramsey voltage modulations :
\begin{align}
t_{1}&=t_{2}=0 \label{eq:t1}\\ 
t_{1}&=t_{2}=t_{0} \label{eq:t2}\\ 
t_{1}&=0 \quad t_{2}=t_{0} \label{eq:t3} \\ 
t_{1}&=t_{0} \quad t_{2}=0. \label{eq:t4}
\end{align}
The noise under the Ramsey voltage modulation can be approximated by the sum of these four main contributions. The equal time contributions \eqref{eq:t1}-\eqref{eq:t2} are the noise due to a single voltage pulse and are therefore trivial. The sum over the contributions from \eqref{eq:t3}-\eqref{eq:t4} shows the Ramsey interference between a particle tunneling at the first pulse ($t=0$) and a particle tunneling at the second pulse ($t=t_0$). 
The resulting expression for the noise is
\begin{multline} \label{eq:S(t_0)}
S(\omega>0,t_0)\sim \frac{1}{4}q^{2}D^{2}\sum_{\eta_{1}\eta_{2}=\pm} 
\\
\bigg[2+2\text{Re}\bigg\{e^{-i\varphi_0}e^{i(\omega-qV_{dc}) t_0}G_L^{\eta_1\eta_2}(t_0) G_R^{\eta_1\eta_2}(t_0)\bigg\} \bigg]
\end{multline}
where we have invoked the time-reversal symmetry of the Green's functions ${G_{R/L}^{\eta\eta'}\left(t\right)=G_{R/L}^{\eta'\eta}\left(-t\right)}$ to express the contribution of \eqref{eq:t3} as the complex conjugate of the contribution of \eqref{eq:t4}.
Eq. \eqref{eq:S(t_0)} is applicable to other quantum Hall states. 

The current noise exhibits a Ramsey interference pattern oscillating with the delay time between the pulses $t_0$ at frequency $\omega-qV_{dc}$. We show below that the charge appearing in this formula is the dressed charge in the fluid meaning the charge of the Laughlin quasiparticles for the case of Fig. 1(a) and the charge of the electron for the case of Fig. 1(b).  The amplitude of the oscillations provides informations on the Green's functions of the tunneling particles.
Thus, measuring the noise as a function of the time delay $t_0$ also allows us  to detect the Green's functions of the particles in the time domain.

For Laughlin states the Green's functions of the fractional quasi-particles is given by \cite{Kim2006}
\begin{equation} \label{eq: G}
G_{L/R,f}^{\eta_1\eta_2}(t)
=\left[\frac{\sin\left(i\chi_{\eta\eta'}\left(t\right)\pi T\left(t\pm\frac{x}{v}\right)\right)}{\pi T\tau}\right]^{-\nu}
\end{equation}
where $\tau$ is a short-time cut-off and the function $\chi_{\eta\eta'}(t)$ is defined as
\begin{equation}\label{eq:chi}
\chi_{\eta\eta'}\left(t\right)\equiv\frac{\eta+\eta'}{2}\text{sgn}\left(t\right)-\frac{\eta-\eta'}{2}.
\end{equation}

Using eq. \eqref{eq:S(t_0)} we find the current noise under the Ramsey voltage modulation
\begin{multline} \label{eq:S_qp}
S_f(\omega>0)\approx2(e\nu)^2D^2\Bigg[1+\cos{\pi\nu}\\
\times\left(\frac{\sinh\left(\pi T t_0\right)}{\pi T\tau}\right)^{-2\nu}\cos\left( (\omega-e\nu V_{dc}) t_0-\varphi_0\right) \Bigg].
\end{multline}

In the high temperature limit $T\gg t_0^{-1}$ the amplitude of the Ramsey oscillations decay exponentially as $\exp\{-2\pi\nu T t_0\}$ whereas in the low temperature limit $T\ll t_0^{-1}$ the amplitude is a power low $\left(t_0\over\tau \right)^{-2\nu}$. The crossover between these two regimes can be observed by scanning the delay time $t_0$.

The decay of the Ramsey interference is not a result of a finite lifetime of the excitations. Actually, the fractional quasi-particle $\psi_f^\dagger(q)$ is an eigen-state of the Hamiltonian with energy $\epsilon=vq$ where $v$ is the edge (magnetoplasmon) velocity \cite{Lederer2000}. The decay of the Ramsey interference is a measure of the envelope of the wave-function of the fractional quasi-particle in time (and space by duality).

Next, we find the noise when the tunneling particles are electrons and show the difference between the two cases.
The Green's functions of the  electrons is equivalent to Eq. \eqref{eq: G} under the exchange $\nu\rightarrow\frac{1}{\nu}$.
Therefore noise of the electron current under the Ramsey voltage modulation is given by
\begin{multline} \label{eq:S_e}
S_e(\omega>0)\approx2e^2D^2\Bigg[1+\cos{\frac{\pi}{\nu}}\\
\times\left(\frac{\sinh\left(\pi T t_0\right)}{\pi T\tau}\right)^{-\frac{2}{\nu}}\cos\left( (\omega-e V_{dc}) t_0-\varphi_0\right) \Bigg].
\end{multline}
For the integer quantum Hall situation with $\nu=1$, the quasi-particles can be identified to the electrons and Eqs. \eqref{eq:S_qp} and \eqref{eq:S_e} coincide.

Comparing the fractional quasi-particles' noise (Eq. \eqref{eq:S_qp}) and  the electrons' noise (Eq. \eqref{eq:S_e}) reveals that the fractionalization manifests itself in the Ramsey interference pattern in three different places:
1) The prefactor $(qD)^2$. This is a result of the definition of the current. The fractional charges measured from Schottky noise measures this prefactor. \cite{goldman1995,Picc1997,Saminadayar1997,dolev2008} 
2) The Green's function of the tunneling particles. \cite{radu2008,milliken1996,Chang1996} 
3) The dynamical phase factor $qV_{dc}$. The formation of interferences in time allows one to measure the charge conjugated to the electric field.
We have shown how the nature of the tunneling particles can be detected by the Ramsey interference, including their charge and Green's function.

The Ramsey signal is a result of interferences between two tunneling events at the same QPC separated by the time $t_0$, while in the Mach-Zehnder geometry \cite{Markus}, the interference is a result of interference between tunneling events at two different QPCs.
The time delay $t_0$ between the voltage pulses can be scanned in a continuous manner in contrast to the length $L$ of the Mach-Zehnder interferometer making the Ramsey interferometry advantageous for measuring the Green's function.
The phase detected by the Mach Zehnder interference is determined by the magnetic flux enclosed by the two arms of the interferometer, while the Ramsey interference is determined by the dynamical phase accumulated by the particles between the two pulses $qV_{dc}t_0$. 

Mach-Zehnder interferometers can also probe the charge fractionalization \cite{Hur2005}, and the statistics of quasi-particles \cite{Gefen2006,Kitaev2006} since they detect the phase acquired by an edge quasi-particle circling the bulk quasi-particles. In practice, however, the number of quasi-particles in the bulk is diffcult to control. Ramsey interferometry involves only a single QPC, hence it does not depend on the bulk quasi-particles and cannot detect exchange statistics. Another type of time-interferometry experiment is the electronic Hong-Ou-Mandel analog which provides information on the time-shape of the electronic wave packet and its decoherence. \cite{dubois2013,Glattli2013,Marguerite2016}  The Hong-Ou-Mandel geometry involves interferences between two different (although perhaps indistinguishable) particles whereas the Ramsey interference involves a single particle. 

Next, we address momentum-resolved tunneling in which the tunnel junction is long compared to the Fermi wavelength \cite{Auslaender2005,Fiete2005,Steinberg2006,steinberg2008,Berg2009}. The tunneling Hamiltonian reads
\begin{equation}
H_{T}=\frac{D}{2\pi a}\int_{0}^{L}dx \psi_{L}^{\dagger}\left(x,t\right)\psi_{R}\left(x,t\right)e^{iq_{B}x+i\varphi\left(t\right)}+h.c.
\end{equation}
where $L$ is the length of the tunnel junction and $a=v\tau$ is a short scale cut-off. In Fig. 1(b), the tunneling electron acquires a momentum boost $q_{B}=qBd/\hbar$ as a result of the magnetic field $B$ applied perpendicular to the plane, $d$ is the distance between the counter propagating modes. The tunneling current is thus given by
\begin{equation}
I=i\frac{qD}{2\pi a}\int_{0}^{L}dx \psi_{L}^{\dagger}\left(x,t\right)\psi_{R}\left(x,t\right)e^{iq_{B}x+i\varphi\left(t\right)}+h.c.
\end{equation}

Under the Ramsey voltage modulation, to second order in $D$, the noise takes the form
\begin{multline} \label{eq:S_mom1}
S\left(\omega>0\right)	\approx \\
\frac{q^{2}D^{2}L}{2\left(2\pi a\right)^{2}}\sum_{\eta_{1}\eta_{2}}\int_{-L}^{L}dx 
\Bigg[G_{L}^{\eta_{1}\eta_{2}}\left(x,0\right)G_{R}^{\eta_{1}\eta_{2}}\left(x,0\right)e^{-iq_{B}x} \\
	+\text{Re}\left\{ e^{-i\varphi_{0}+i\left(\omega-qV_{dc}\right)t_{0}} G_{L}^{\eta_{1}\eta_{2}}\left(x,t_{0}\right)G_{R}^{\eta_{1}\eta_{2}}\left(x,t_{0}\right)e^{-iq_{B}x} \right\} \Bigg].
\end{multline}

We first assume that the tunneling particles are electrons. The same reasoning will be then applied to the Laughlin quasiparticles in the case of a long tunneling region through the bulk. 
Without loss of generality we take $q_{B}>0$, thus we close the integral contour in Eq. \eqref{eq:S_mom1} in the lower half plane. The poles of $G^{++}$ and $G^{-+}$ are in the upper half plane therefore they do dot contribute to the noise. The main contributions to the integral over $G^{--}$ and $G^{+-}$ come from the divergence at $x=-vt_{0}$ so
\begin{multline}
\int_{-L}^{L}dxG_{L}^{--}\left(x,t_{0}\right)G_{R}^{--}\left(x,t_{0}\right)e^{-iq_{B}x}	\\
=\int_{-L}^{L}dxG_{L}^{+-}\left(x,t_{0}\right)G_{R}^{+-}\left(x,t_{0}\right)e^{-iq_{B}x} \\
	\approx\left[-\frac{\sinh\left(2\pi Tt_{0}\right)}{\pi T\tau}\right]^{-\nu}e^{iq_{B}vt_{0}}.
\end{multline}
In order to find the prefactor we compare it with the exact result for free electrons ($\nu=1$) at zero temperature
\begin{equation}
\int_{-L}^{L}dxG_{L}^{--}\left(x,t_{0}\right)G_{R}^{--}\left(x,t_{0}\right)e^{-iq_{B}x}	=i\frac{\pi v\tau^{2}}{t_{0}}e^{iq_{B}vt_{0}}.
\end{equation}
Hence, we conclude that
\begin{multline}
\int_{-L}^{L}dxG_{L}^{--}\left(x,t_{0}\right)G_{R}^{--}\left(x,t_{0}\right)e^{-iq_{B}x}	\\
= i2\pi u\tau\left[\frac{\sinh\left(2\pi Tt_{0}\right)}{\pi T\tau}\right]^{-\nu}e^{iq_{B}vt_{0}}.
\end{multline}
Finally, the noise under the Ramsey voltage modulation is given by
\begin{multline}
S\left(\omega>0\right)	\approx q^{2}D^{2}\frac{L}{2\pi a} \times \\
 \left(1+\left[\frac{\sinh\left(2\pi Tt_{0}\right)}{\pi T\tau}\right]^{-\nu}\sin\left(\left(\omega-qV_{dc}+q_{B}v\right)t_{0}-\varphi_{0}\right)\right).
\end{multline}
Here, the frequency of the Ramsey oscillations depends also on the momentum boost $q_B$ and on the velocity $v$. 
The amplitude of the Ramsey signal behaves as in  the case of a point-like tunnel junction Eq. \eqref{eq:S_qp}. 

For fractional quasi-particles, the above expression remains valid under the replacement $\nu\rightarrow \nu^{-1}$.

As an example of the ability of Ramsey interferometry to distinguish between different theories, we consider a quantum Hall bar at filling factor $\frac{5}{2}$, which is presumed to support non-Abelian excitations and consequently is of major interest for topological quantum computation \cite{Nayak2008Rev}, and we compare between two of its candidates: the Pfaffian state \cite{MooreRead91}, and its particle-hole conjugate, the anti-Pfaffian state \cite{Lee2007,Levin2007}. 
 The most relevant tunneling process of quasi-particles at a QPC is of quasi-particles with charge $q=\frac{e}{4}$ described by the operator \cite{Levin2007,Fendley2007,Bishara2008}
\begin{equation}
\psi_{P/AP}\propto\sigma e^{ i\frac{1}{2}\left(\phi_{c}\pm\alpha\phi_n\right)}
\end{equation}
where $\sigma$ is an Ising spin field, $\phi_c$ is a charged bosonic field and $\phi_n$ is a neutral bosonic field. Here  $P/AP$ denote the Paffian and anti-Paffian states with  $\alpha=0$ in the Pfaffian state and $\alpha=1$ in the anti-Pfaffian state.

Using the corresponding Green's functions \cite{Levin2007,Fendley2007,Carrega2012,Bishara2008,ginsparg1989}  in Eq. \eqref{eq:S(t_0)} we find the current noise under the Ramsey voltage modulation 
\begin{multline}
S(\omega>0,t_0)\sim2\left(\frac{e}{4}\right)^{2}D^{2}
\bigg[1+
\cos\left(\pi\frac{1+\alpha^2}{4}\right)\\
\times\left[\frac{\sinh\left(\pi Tt_0\right)}{\pi T\tau}\right]^{-\frac{1}{2}-\frac{\alpha^2}{2}}
\cos\left(\omega-\frac{e}{4}V_{dc}-\varphi_0^*\right) \bigg],
\end{multline}
where $\varphi_0^*\equiv\varphi_0+\frac{\pi}{16}\left(3+8\alpha^2\right)$.
Thus, Ramsey interferometry can differentiate between the Pfaffian and anti-Pfaffian states.
Recent experiments support the particle-hole Pfaffian state for the description of the state at $\nu=\frac{5}{2}$ \cite{Son2015,Zucker2016,banerjee2018}. The particle-hole Pfaffian state is similar to the Pfaffian state, but the Majorana field propagates in the opposite direction to the charged boson. Therefore, Ramsey interferometry can not distinguish between the Pfaffian and the particle-hole Pfaffian states.

To conclude, we suggest to probe the time dependent dynamical phase of fractional quasi-particles as well as their Green's function in time by generalizing Ramsey interferometry to quantum Hall edge states. This probe is then useful to characterize topological states of matter and  light \cite{lightreview}, and could be generalized to superconductors through the time dynamics of Bogoliubov quasiparticles. 

Acknowledgements:  This work was supported by the Labex PALM Paris-Saclay ANR-10-LABX-0039 and by the
Deutsche Forschungsgemeinschaft (DFG, German Research
Foundation) via Research Unit FOR 2414 under
project number 277974659. This work has also benefitted from discussions at CIFAR meetings in Canada and at the Centre de Recherches in Math\' ematiques in Montreal.

\bibliographystyle{apsrev4-1}
\bibliography{Ramsey_FQH_ref.bib}

\end{document}